\documentclass[aip,jcp,twocolumn,reprint,longbibliography]{revtex4-1}
\usepackage[usenames, dvipsnames]{color}
\usepackage{graphicx}
\usepackage{color}
\usepackage{epstopdf}
\usepackage{epsfig}
\usepackage{bm}
\usepackage{xcolor}
\usepackage{amsmath}
\usepackage{amsfonts}
\usepackage{subfigure}
\usepackage{verbatim}
\usepackage[version=3]{mhchem}
\usepackage[ansinew]{inputenc}
\usepackage{epstopdf}
\usepackage{scalerel}
\usepackage{multirow}
\usepackage{booktabs}
\usepackage[para,flushleft]{threeparttable}

\newcommand{\avg}[1]{\left<#1\right>}
\newcommand{\len}[1]{\left|#1\right|}

\newcommand{\para}[1]{\left(#1\right)}

\newcommand{\kT}{\ensuremath{k_{\rm B}T}}

\newcommand{\Lb}{\ensuremath{\mathcal{L}}}

\newcommand{\Rbar}{\overline{\textbf{R}}}

\newcommand{\EE}{\textit{E}}

\begin{document}


\title{Dielectric Saturation in Water from a Long Range Machine Learning Model}


\author{Harender S. Dhattarwal}
\affiliation{Department of Chemistry and Chemical Biology, Rutgers University, Piscataway, NJ 08854}
\author{Ang Gao}
\affiliation{Department of Physics, Beijing University of Posts and Telecommunications, 100876, Beijing, China}
\author{Richard C. Remsing}
\email[]{rick.remsing@rutgers.edu}
\affiliation{Department of Chemistry and Chemical Biology, Rutgers University, Piscataway, NJ 08854}



\begin{abstract}
Machine learning-based neural network potentials have the ability to provide ab initio-level predictions while reaching large length and time scales often limited to empirical force fields. 
Traditionally, neural network potentials rely on a local description of atomic environments to achieve this scalability.
These local descriptions result in short range models that neglect long range interactions necessary for processes like dielectric screening in polar liquids. 
Several approaches to including long range electrostatic interactions within neural network models have appeared recently,
and here we investigate the transferability of one such model, the self consistent neural network (SCFNN), which focuses on learning the physics associated with long range response. 
By learning the essential physics, one can expect that such a neural network model should exhibit at least partial transferability. 
We illustrate this transferability by modeling dielectric saturation in a SCFNN model of water. 
We show that the SCFNN model can predict non-linear response at high electric fields, including saturation of the dielectric constant,
without training the model on these high field strengths and the resulting liquid configurations.
We then use these simulations to examine the nuclear and electronic structure changes underlying dielectric saturation. 
Our results suggest that neural network models can exhibit transferability beyond the linear response regime
and make genuine predictions when the relevant physics is properly learned.
\end{abstract}
\maketitle


\section{Introduction}
\label{sec:introduction}

Ab initio molecular dynamics simulations enable computational predictions of
interatomic interactions and chemical reactivity~\cite{Car1985,Sun:NatChem:2016,Chen2017,Tuckerman1996,marx2009,Kresse1993,szabo2012,Dhattarwal2022,Sung2022}, 
but the expense of performing electronic structure calculations limits their use to small system sizes and short time scales. 
Empirical interaction potentials, or force fields, are regularly used to model large systems and long time scales~\cite{Karplus1990,HANSSON2002,Riniker2018,Venable2019,Brooks2021},
but it is difficult to include processes like chemical reactions and electronic polarization in these classical models~\cite{Leven2021,Shi2021,Jing2019,Bedrov2019}. 
To bridge the gap between ab initio and force field-based simulations,
machine learning-based neural network models are being developed to achieve ab initio accuracy
at a fraction of the computational cost~\cite{Behler2007,Zhang2018,Hansen2015,Artrith2011}.
Neural network models learn interatomic interactions from ab initio calculations to enable
efficient sampling of potential energy surfaces~\cite{Blank1995,Rowe2020,Frank2019,Behler2016,Yao2022}.
By learning ab initio-level interactions, the resulting neural network models
are able to predict electronic polarization effects and bond breakage and formation
in large systems and on larger timescales than accessible in current ab initio simulations~\cite{Behler2017,Xu2021,Gerwen2022,Vandermause2022,Cools2022,Manzhos2021,Valeria2022,debenedetti2020}.
To reach these large scales, neural network models often operate under the assumption of locality,
in which interatomic interactions are determined by atomic arrangements within a spherical region
typically on the scale of 1~nm or less~\cite{Behler2016,Behler2007,Behler2021}.
These methods have created profound insights into chemical systems and revolutionized
molecular simulations~\cite{Behler2016,Manzhos2021,kapil2022first,gartner2022liquid}.
However, this assumption of locality results in short range models that lack a description
of long range electrostatics~\cite{Yue2021,Limmer2021,Behler2021,Remsing2022}. 
As a result, several approaches to modeling long range interactions in neural network models
have begun to appear~\cite{Artrith2011,Grisafi2019,Ko2021,Zhang2022,Yao2018,Remsing2022,pagotto2022predicting}.
The recently-developed self-consistent field neural network (SCFNN)
separately learns short and long range interactions through two coupled neural networks~\cite{Remsing2022}.
This enables the SCFNN to learn long range response,
while including the impact of long range effects on short range structure
and interactions through a rapidly converging self-consistent loop.
The resulting SCFNN model can accurately describe dielectric screening and the response
of a system to external electrostatic fields. 
Importantly, the SCFNN uses a physically-meaningful separation of interactions
into short and long range interactions~\cite{LMFDeriv,Remsing:PNAS:2016,SSM} and focuses on learning the physics underlying the response to long range fields.
Consequently, the SCFNN is partially transferable to environments for which it was not trained.
Training the SCFNN involves computing energies and forces for configurations of a system,
as well as the energies and forces of those same nuclear configurations in the presence of electric fields. 
Despite this training being performed on configurations and fields in the linear response regime,
the SCFNN model makes no assumptions of linear response.
As a result, the SCFNN should be able to describe non-linear response.
Here, we illustrate the transferability of the SCFNN model to the non-linear response regime
by modeling dielectric saturation in liquid water.
When an external electric field is applied to a polar liquid,
the liquid will respond to screen the field through changes in electronic and nuclear structure~\cite{onsager1936electric,kirkwood1939dielectric,frohlich1949theory,Bottcher,Zangwill,chandler1977dielectric,de1986computer,Yeh1999,zhang2020modelling,Seyedi2019}.
This response consists mainly of changes in the orientational nuclear structure of the liquid,
such that molecules reorient their dipoles in the direction of the field.
For small fields, the liquid's response is linear and determined by its static dielectric constant~\cite{onsager1936electric,kirkwood1939dielectric,frohlich1949theory,Bottcher,purcell2013electricity,Zangwill,chandler1977dielectric,TheorySimpLiqs,Cox2019,Cox2020}.
However, as the field is increased in magnitude, the liquid does not continue to respond linearly;
there is a limit to the amount of dipolar reorientation that can occur.
As the maximum value of dipolar reorientation is approached, the liquid responds non-linearly to
the external electric field~\cite{debye1929polar,Booth1951,booth1955dielectric,Sutmann1998,Alper1990,Sandberg2002,Sprik2016,Yeh1999,Matyushov2015,matyushov2018nonlinear,richert2017nonlinear,kotodziej1975high,jeanmairet2019study,willard2009water}.
Consequently, dipolar fluctuations are damped, and there is a reduction in the dielectric constant --- dielectric saturation.
The non-linear response to external electric fields underlying dielectric saturation
can be used as a test on the transferability of the SCFNN model.
Here, we show that the SCFNN model can describe dielectric saturation in water without
including configurations consistent with this effect in the training data.
After demonstrating that the SCFNN is transferable to the non-linear response regime,
we examine classic theories and quantify the structural underpinnings of dielectric saturation,
including electronic polarization effects.
We conclude with a discussion of our results in the context of neural network models.
%


\section{Simulation Details}
\label{sec:method}

We modeled the dielectric response of water to the applied homogeneous electrostatic fields using molecular dynamics simulations. 
Eight different electric fields, 0.005, 0.01, 0.015, 0.02, 0.05, 0.10, 0.20, and 0.28~V/\AA, were applied to a cubic box of
1000 water molecules with dimension 31.2~\AA. 
A total of 56 independent trajectories were sampled for the three fields less than 0.02~V/\AA, and a total of 28 independent trajectories were sampled for the rest.
Each system was equilibrated for 50~ps at 300~K in the NVT ensemble,
with the temperature maintained using the Berendsen thermostat~\cite{Berendsen1984}.
The equations of motion were integrated using a time step of 0.5~fs. 
The systems were periodic in all three directions. 
The last 25~ps of all trajectories were used for analysis.
The SCFNN was trained following previous work~\cite{Remsing2022}.
Training the SCFNN includes computing energies and forces for configurations of water with fixed nuclear positions
in the absence and presence of electric fields.
The original SCFNN model of water used electric field of magnitude 0.1 and 0.2~V/\AA \ to learn the
long range electronic response of water to applied fields~\cite{Remsing2022}.
We refer to this model as SCFNN(HF) to indicate that it was trained at high field strengths.
In this study, we trained another SCFNN model using smaller electric fields
of magnitude 0.005 and 0.01~V\AA, 
which corresponds to the linear polarization regime for both electronic and nuclear response, based on previous studies~\cite{Sprik2016,Cox2019}. 
We refer to this as the SCFNN model throughout this work.
The network architecture of both models is the same.
We use a Behler-Parrinello style network for the short range part of the SCFNN~\cite{Behler2007,morawietz2016van,Remsing2022},
though we expect that many established approaches could readily be used for the short system.
The DFT calculations for training followed previous work~\cite{Dellago,marsalek2017quantum,Remsing2022} and
used the CP2K program~\cite{VandeVondele2005,Kuhne2020} with
the revised Perdew-Burke-Ernzerhof hybrid exchange-correlation functional with 25\% exact exchange (revPBE0)~\cite{Adamo1999,Zhang1998,Goeigk2011} and
k-point sampling at the $\Gamma$-point only. 
Goedecker-Teter-Hutter (GTH) pseudopotentials~\cite{Goedecker1996} were used with TZV2P basis sets~\cite{VandeVondele2007}.
DFT-D3 dispersion corrections were used to include long ranged van der Waals interactions~\cite{Grimme2010}.
The calculations used an energy cutoff of 400~Ry and 60~Ry for the reference grid (keyword REL\_CUTOFF). 
Maximally localized Wannier function centers were obtained by minimizing the spread of MLWFs within CP2K~\cite{Marzari2012,Berghold2000}. 
Configurations for the training data were taken from previous work~\cite{Dellago}.
%


\begin{figure}[tb]
\begin{center}
\includegraphics[width=3in]{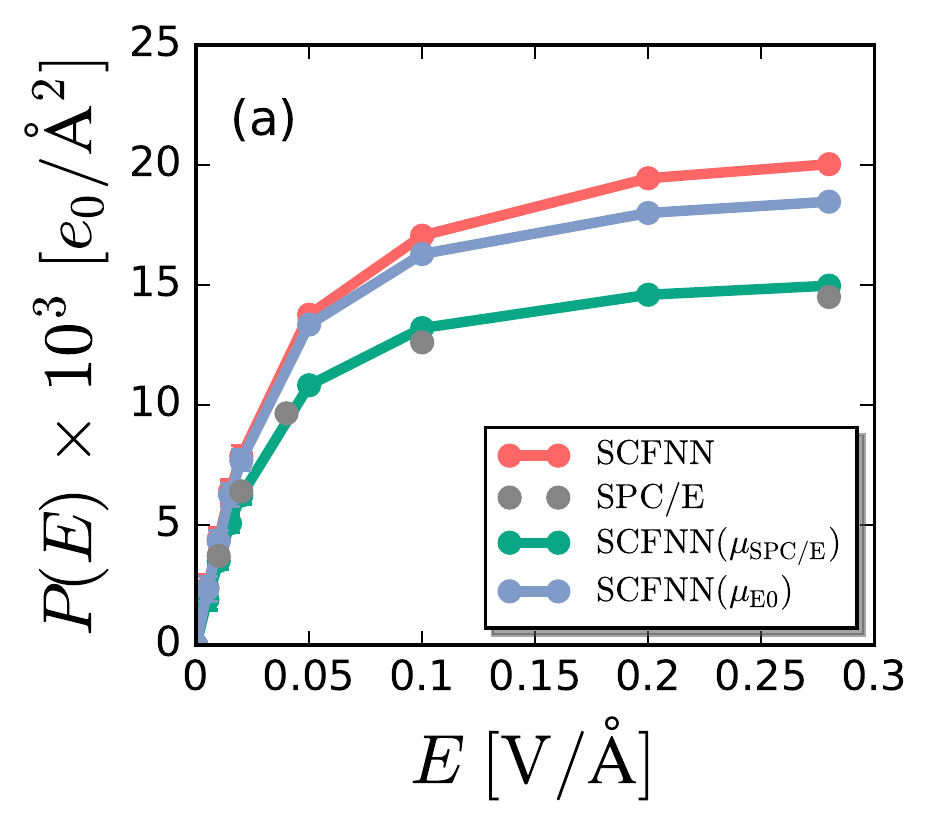}
\includegraphics[width=3in]{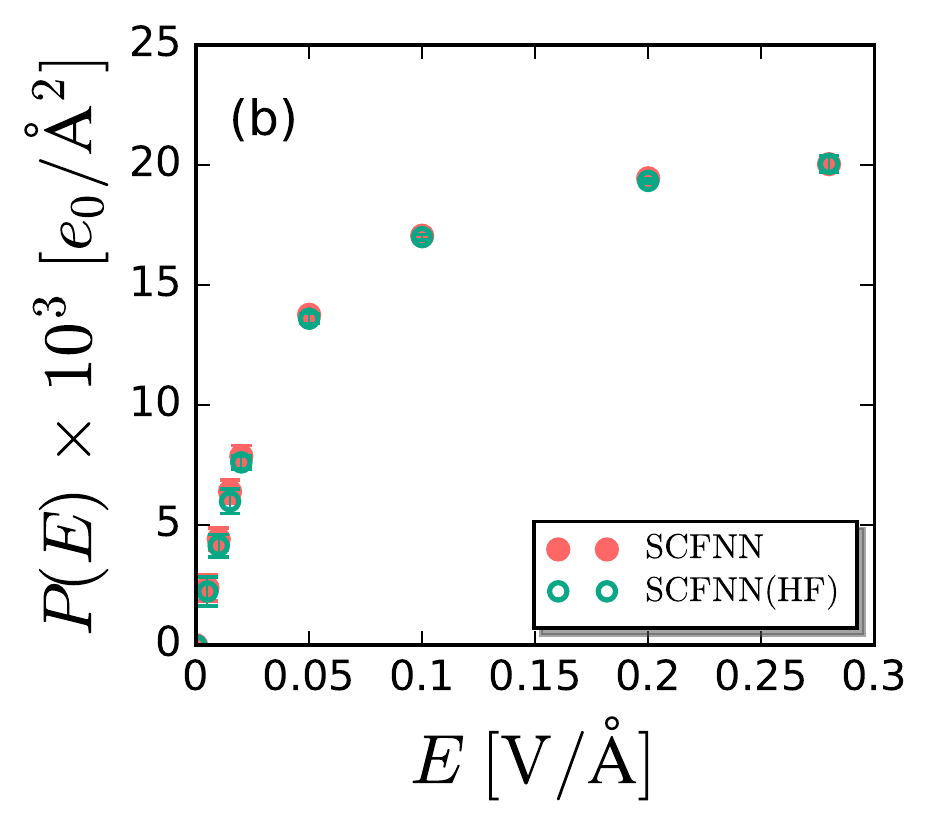}
\caption{\label{fig:PvE} (a) Polarization as a function of electric field strength for the SCFNN and SPC/E
water models; SPC/E data is from previous work~\cite{Sprik2016}.
Also shown is the polarization computed using the structure predicted by the SCFNN model
but with the SPC/E dipole moment, SCFNN($\mu_{\rm SPC/E}$),
and with the value of the SCFNN dipole moment at zero field, SCFNN($\mu_0$).
(b) Comparison of the polarization as a function of electric field strength for the SCFNN and SCFNN(HF) models.
Error bars indicate one standard deviation.}
\end{center}
\end{figure}

\section{Results and Discussion}
\label{sec:rd}

\begin{figure}[tb]
\begin{center}
\includegraphics[width=3in]{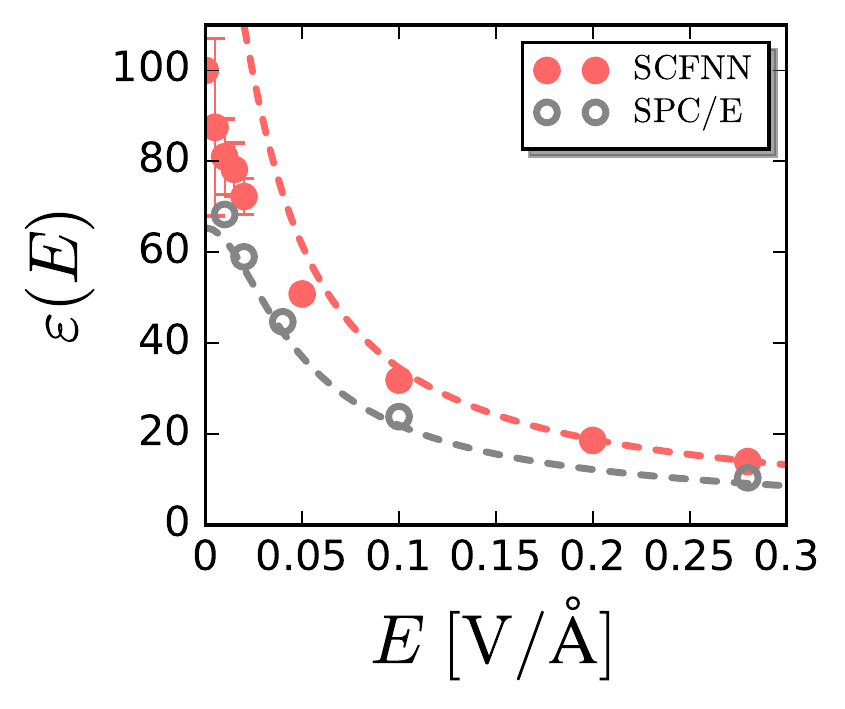}
\caption{\label{fig:EpsvE} The electric field-dependence of water's dielectric constant, $\varepsilon(E)$.
Dashed lines correspond to predictions from the Kirkwood-Booth Equation~\ref{eq:booth},
using $\varepsilon_\infty=1.65$ for SCFNN and $\varepsilon_\infty=1$ for SPC/E.
Error bars indicate one standard deviation.}
\end{center}
\end{figure}

To examine the nonlinear response of water to applied electric fields,
we apply uniform fields of magnitude \EE~to water and compute the 
polarization induced in the system according to~\cite{Cox2019}
\begin{equation}
\centering
\label{eq:PvE}
    P(E)  = \avg{P(\Rbar)}_{E}  = \avg{ \frac{1}{ V }  \sum_{i=1}^N \mu_{i}(\Rbar)}_E.
\end{equation}
Here, $\avg{\cdots}_E$ indicates an ensemble average over configurations $\Rbar$ in the presence
of a field of strength \EE, such that $P(\Rbar)$ is the instantaneous polarization of a single configuration $\Rbar$,
$V$ is the volume of the simulation cell containing $N$ water molecules, and $\mu_i(\Rbar)$ is the
instantaneous dipole moment of molecule $i$.
The polarization is linear at low field strength, but becomes non-linear beyond approximately 0.02~V/\AA, Fig.~\ref{fig:PvE}.
For large enough fields, $P(E)$ begins to plateau, indicative of dielectric saturation~\cite{Alper1990,Yeh1999,Sutmann1998,Sandberg2002,Cox2020,Sprik2016}.
This non-linear behavior is well-described by the SCFNN model despite it not being trained in this regime. 
We also compare the response of the SCFNN to that of the SPC/E water model obtained in previous work~\cite{Sprik2016}.
The SPC/E model displays similar behavior although its polarization in the nonlinear regime is significantly smaller
than that of the SCFNN model.
Classic treatments of dielectric response treat water as a collection of independent dipoles, such that the
orientation and magnitude of the molecular dipole moment determines the polarization. 
From this perspective, the SPC/E and SCFNN models should have different polarization values because
their average dipole moments differ: 2.9~D for SCFNN in zero field and 2.35~D for SPC/E.
To further this comparison, we computed the polarization of SCFNN configurations assuming
that the magnitude of every dipole is the same as that of SPC/E. 
The resulting polarization, indicated by SCFNN($\mu_{\rm SPC/E}$) in Fig.~\ref{fig:PvE}a,
is very similar to that of the SPC/E model. 
This agreement suggests that the change in the orientation of molecular dipoles induced by the field
is similar in the SCFNN and SPC/E models. 
Training the SCFNN model used here used lower field strengths than the original SCFNN model~\cite{Remsing2022}. 
To ensure that the model is robust, we performed the same simulations using the original model, termed SCFNN(HF) to indicate
that the fields used in training this model, 0.1~and 0.2~V/\AA, were higher than those used to train what we refer to as the SCFNN model, 0.005~and 0.01~V/\AA.
The polarization predicted by both models is essentially identical, Fig.~\ref{fig:PvE}b.
This excellent agreement further suggests that the SCFNN model is learning the underlying physics responsible for screening and, as a result, this long range part
of the neural network model is transferable. 
The remaining results are shown only for SCFNN unless indicated otherwise; the SCFNN(HF) yields essentially identical results. 
We can use our predictions of the polarization to estimate the
field-dependent dielectric constant, $\varepsilon(E)$, from the derivative of the polarization with respect to the field,
according to~\cite{frohlich1949theory,Bottcher,purcell2013electricity,Zangwill,Kusalik1994,Sprik2016}
\begin{equation}
\centering
\label{eq:EpsvsE}
    \varepsilon(E)  =  1 + \frac{4 \pi P(E)}{ E }.
\end{equation}
As field strength is increased, $\varepsilon(E)$ monotonically decreases and begins to plateau at high field strengths (Fig.~\ref{fig:EpsvE}).
This is a clear indication of dielectric saturation at high fields. 
The dielectric constant of the SCFNN model is larger than that of SPC/E but follows the same general trend.
%

%
At high field strengths, the dielectric constant can
be described by the Kirkwood-Booth equation~\cite{kirkwood1939dielectric,frohlich1948general,Booth1951}
\begin{equation}
\centering
\label{eq:booth}
     \varepsilon(E)  = \varepsilon_\infty + \frac{7 N (\varepsilon_\infty + 2)}{3 V \epsilon_{0} \sqrt{73}} \frac{\mu_0}{E} \Lb\para{ \frac{\sqrt{73} (\varepsilon_\infty + 2)}{6 k_{\rm B} T} \frac{\mu_0}{E}},
\end{equation}
where $\varepsilon_\infty$ is the high frequency dielectric constant,
\begin{equation}
\mu_E=\avg{\frac{1}{N}\sum_{i=1}^N\len{\mu_i(\Rbar)}}_E
\end{equation}
is the average magnitude of the water dipole in the presence of an electric field of magnitude $E$,
such that $\mu_0=\mu_{E=0}$,
and $\Lb(x)=\coth(x)-1/x$ is the Langevin function~\cite{langevin1905magnetisme}. 
The Kirkwood-Booth equation neglects solvent-solvent correlations
beyond the first coordination shell and
assumes that the first coordination shell is independent of field strength
and described by the Bernal-Fowler model of water~\cite{bernal1933theory}.
Despite these simplifications, Eq.~\ref{eq:booth} was shown to provide a reasonable description of dielectric saturation.
In part, the accuracy of the model was achieved by adjusting the dipole moment of water from that in the gas phase to $\mu_0=2.1$~D,
and fitting the value of the dipole moment in this way compensates for errors arising from the above approximations~\cite{Booth1951}. 
We find that the Kirkwood-Booth model with $\mu_0$ determined from simulations qualitatively produces dielectric saturation,
is quantitatively accurate in the high field regime for the SCFNN model, and describes the dielectric constant nearly everywhere for the SPC/E model, Fig.~\ref{fig:EpsvE}.

\begin{figure}[tb]
\begin{center}
\includegraphics[width=3in]{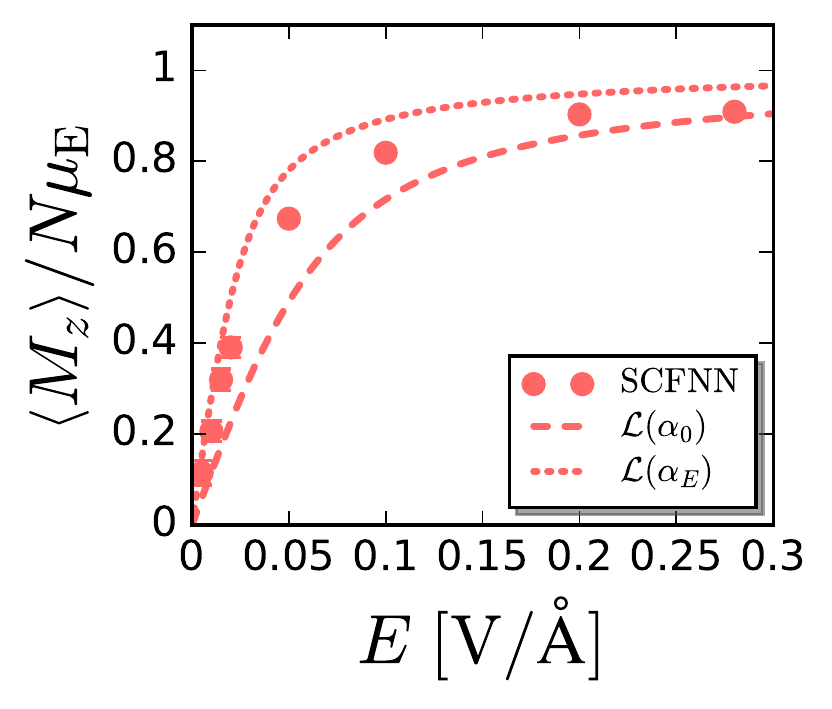}
\caption{\label{fig:Lang} Total dipole moment of the system relative to the maximum value,
equivalent to $\avg{\cos\theta_z}$, where $\theta_z$ is the angle between the molecular dipole moment vector and the direction of the field.
Dashed and dotted lines correspond to predictions from the Langevin function, Eq.~\ref{eq:Lang}, with $\alpha_0$ and $\alpha_E$, respectively.
Error bars indicate one standard deviation.}
\end{center}
\end{figure}

%
In a similar manner, the polarization can be estimated using the Langevin function~\cite{debye1929polar,Booth1951,langevin1905magnetisme},
\begin{equation}
\centering
\label{eq:Lang}
 \frac{\avg{M_z}_E}{ N\mu_E } =  \avg{\cos\theta_z}_E \approx \Lb\para{\alpha_0}, 
\end{equation}
where $\avg{M_z}_E$ is the total dipole moment of the system in the presence of the field \EE,
$\theta_z$ is the angle made by the water dipole moment vector and the direction of the field,
$\alpha_E = 3\mu_E/2\kT$, and $\kT$ is the product of Boltzmann's constant and the temperature.
Equation~\ref{eq:Lang} arises from an independent dipole approximation,
wherein each dipolar molecule is embedded in a dielectric medium and intermolecular correlations between the dipoles are ignored~\cite{Booth1951,langevin1905magnetisme}.
While the Langevin function produces the general trend, it underestimates the
value of $\avg{\cos\theta_z}_E$ predicted by the simulations for all but the smallest and largest fields.
Moreover, the agreement with the SCFNN results is deceptive because 
the Langevin function assumes that the magnitude of the dipole moment is independent of field strength,
but the dipole moment can change in the SCFNN model.
The good agreement between SCFNN and the Langevin prediction results from scaling $\avg{M_z}_E$ by $\mu_E$
and not $\mu_0$; scaling by $\mu_0$ can result in values larger than one~\cite{piekara1962dielectric}. 
As a result, one may anticipate that polarization in the SCFNN model includes a non-zero electronic contribution.
%

\begin{figure}[tb]
\begin{center}
\includegraphics[width=3in]{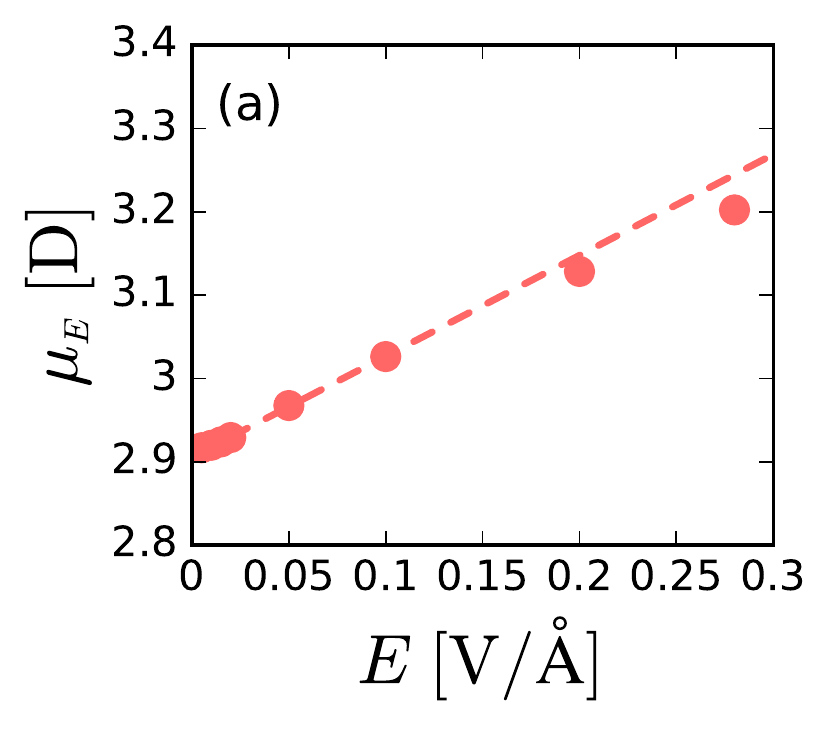}
\includegraphics[width=3in]{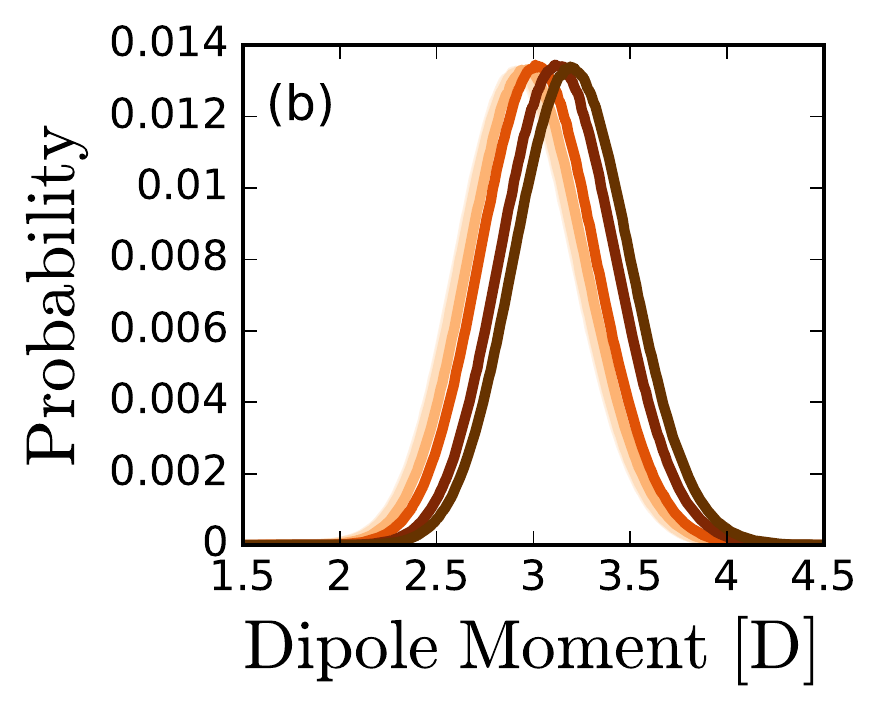}
\caption{\label{fig:Dipoles} (a) Average molecular dipole moment of water as a function of electric field strength.
The dashed line shows a linear fit to the initial increase between $0.02$~and $0.1$~V/\AA.
(b) The probability distribution of the water dipole moment for all field strengths studied here.
The darkness of the lines are proportional to the field strength.}
\end{center}
\end{figure}

\begin{figure*}[tbp]
\begin{center}
\includegraphics[width=5.5in]{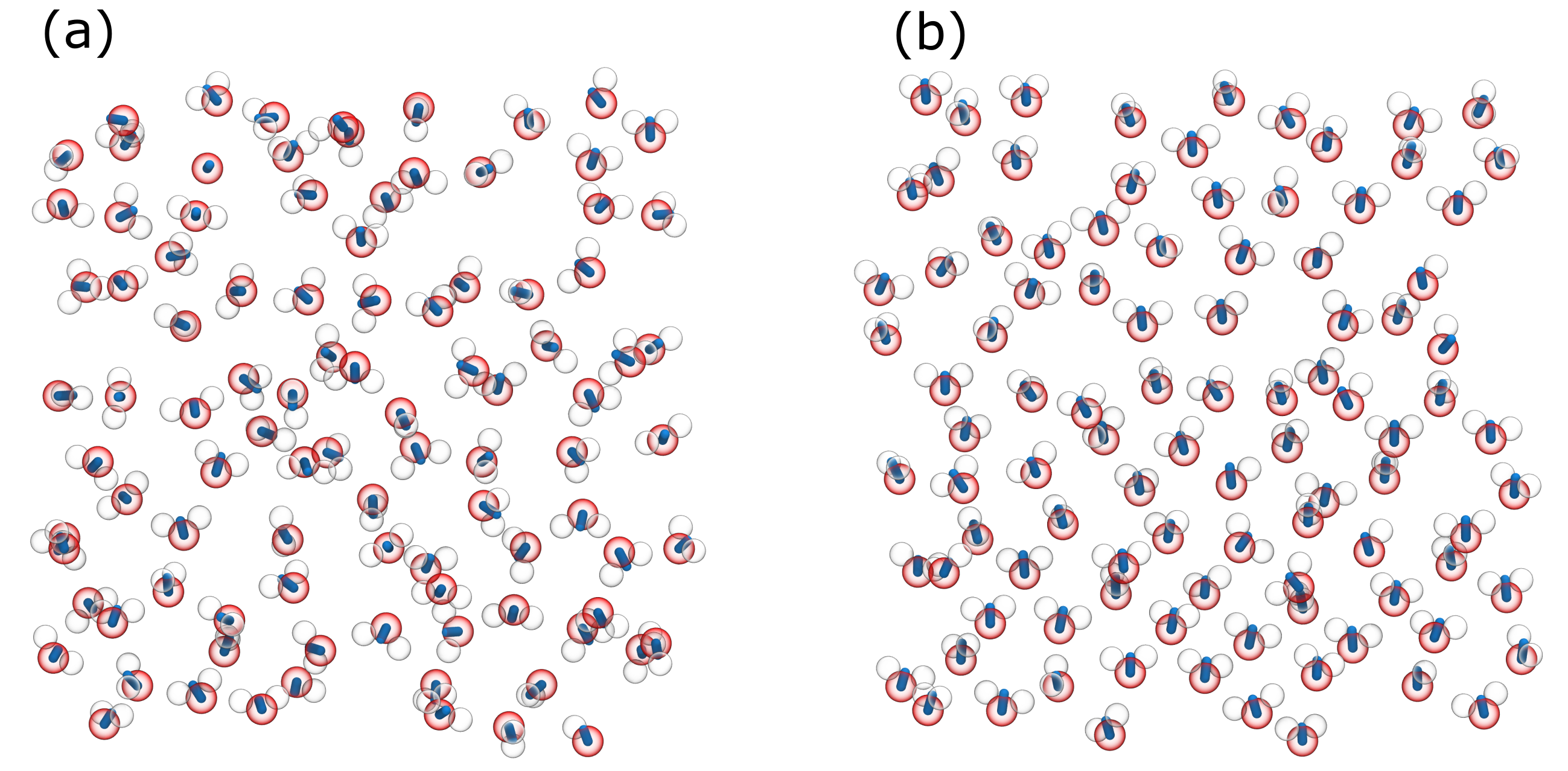}
\caption{\label{fig:DipoleRep} Simulation snapshots illustrating the dipoles (blue lines) of water molecules (red oxygen and white hydrogen)
under constant electric fields of strength (a) 0.005 and (b) 0.28 V/\AA, respectively.}
\end{center}
\end{figure*}

\begin{figure*}[tb]
\begin{center}
\includegraphics[width=0.68\textwidth]{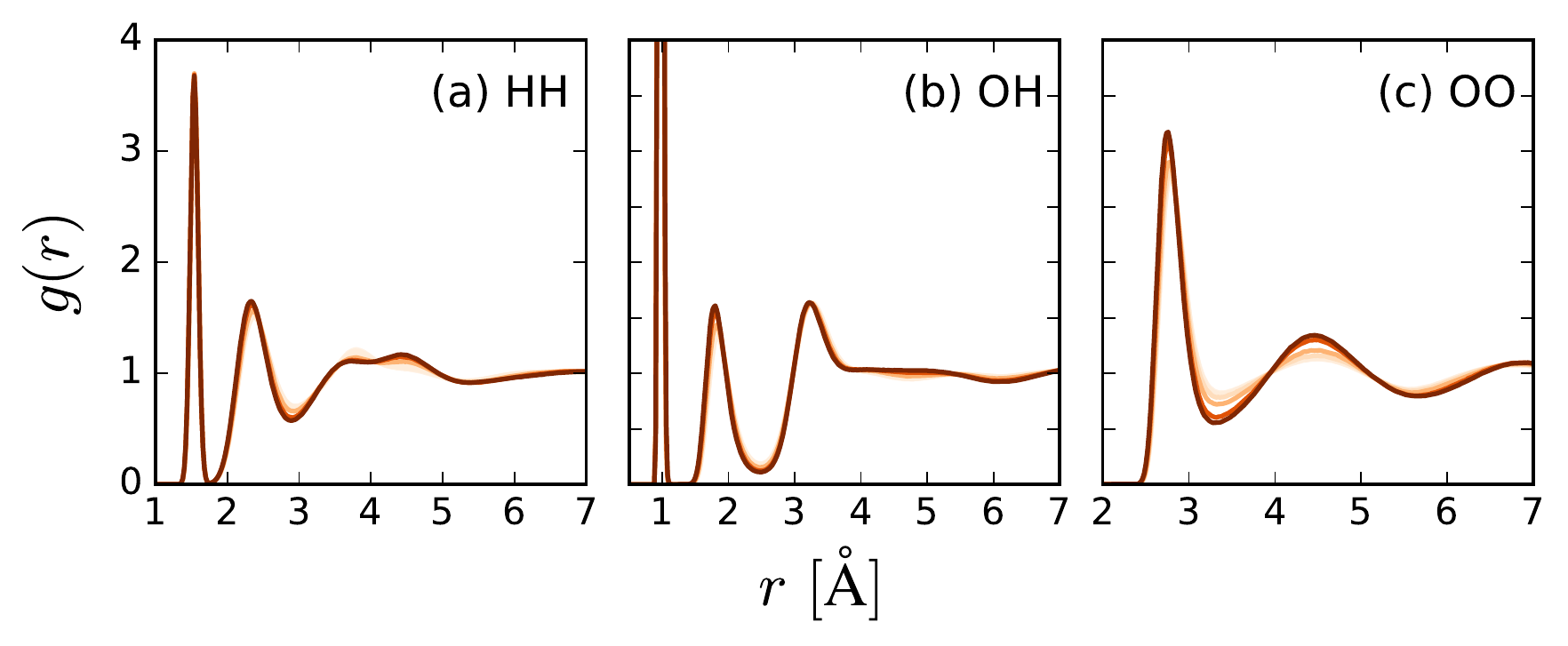}
\includegraphics[width=0.3\textwidth]{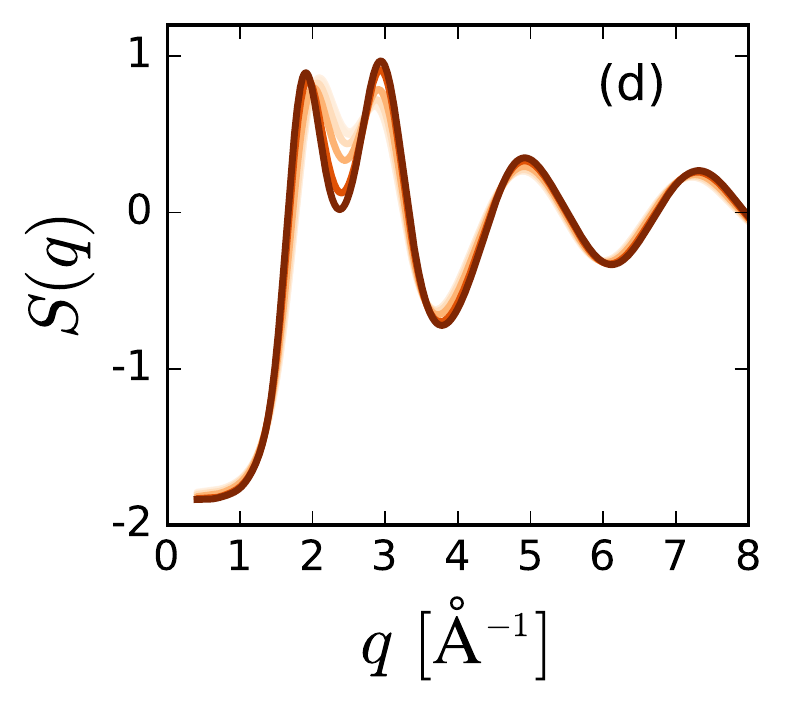}
\caption{\label{fig:RDF} The (a-c) radial distribution functions, $g(r)$, and (d) total X-ray scattering structure factor, $S(q)$, quantifying interatomic correlations in water.
The darkness of the lines are proportional to the field strength.}
\end{center}
\end{figure*}

%
We examine the average molecular dipole moment of water as a function of field, $\mu_E$ (Fig.~\ref{fig:Dipoles}).
At low fields, in the linear regime, the average molecular dipole moment is nearly constant.
At high field strengths, the average molecular dipole moment increases with field strength,
as anticipated from the discussion above.
Unlike the polarization, the increase in the dipole moment does not saturate at high \EE,
although it does increase nonlinearly at the highest field values studied here.
This suggests that the electronic polarization of water is beginning to saturate as well. 
The nature of the fluctuations in water's molecular dipole moment are characterized by
its probability distribution for varying electric field strengths (Fig.~\ref{fig:Dipoles}b). 
We find that the dipole moment fluctuations remain Gaussian for all fields studied,
evidenced by the constant width of the dipole moment probability distributions. 
At low fields, in the linear response regime, the dipole moment distributions are constant with increasing field,
consistent with the constant dipole moment at low fields (Fig.~\ref{fig:Dipoles}a).
At high fields, the only change in the distributions is a shift of their mean,
consistent with the non-linear increase in the dipole moment.
Interestingly, the dipole moment continues to increase at the highest fields studied,
despite the structural changes saturating at high fields. 
This suggests that water can continue to respond to an applied field through electronic polarization
even when the nuclear structure can no longer produce an increase in polarization.
To quantify the contribution of electronic polarization to the total polarization of water at high fields,
we computed the polarization from SCFNN with the magnitude of every dipole moment replaced by the average value at zero field,
$\mu_E\approx\mu_0$.
At low fields, the polarization produced by this approximation is the same as that computed in the simulations (Fig.~\ref{fig:PvE}). 
However, the polarization obtained in this approximation begins to underestimate the actual polarization at high fields. 
The difference between these two curves reflects the contribution of electronic polarization to the total,
which arises from changes in the magnitude of the molecular dipole moment of water.
One can attempt to modify traditional theories like the Langevin function to account for the field dependence of $\mu_E$.
Inserting $\mu_E$ into the Langevin function of Eq.~\ref{eq:Lang}, $\avg{\cos\theta_z}_E\approx \Lb(\alpha_E)$, results in better agreement with the SCFNN model,
as shown in Fig.~\ref{fig:Lang}, where we estimated $\mu_E$ for all fields by fitting the simulation results to a third order polynomial.
The increase in the dipole moment at higher fields increases the estimate of $\avg{\cos\theta_z}_E$,
so much so that the SCFNN results are now slightly overestimated.
This overestimation likely arises from the neglect of correlations in the Langevin function (the independent dipole approximation).
In water, the hydrogen bond network places constraints on the orientations that molecules can adopt,
and so water molecules cannot align with the applied field to the same extent as independent dipoles,
resulting in a lower $\avg{\cos\theta_z}_E$~\cite{piekara1962dielectric}.
While using $\mu_E$ instead of $\mu_0$ may improve the predictions of Eq.~\ref{eq:Lang}, making the same substitution in Eq.~\ref{eq:booth} results in minimal changes
in its estimate of the dielectric constant.
The saturation of the polarization at high field strengths occurs when water molecules maximally orient
their dipoles along the field direction, as shown in Fig.~\ref{fig:DipoleRep}.
At low field strengths, the dipolar structure of the liquid is disordered (Fig.~\ref{fig:DipoleRep}a).
This is consistent with water exhibiting a linear response at low fields, because the liquid structure
is essentially the same as that at zero field (see below for more details). 
In contrast, at high field strength, water dipoles preferentially align in one direction (Fig.~\ref{fig:DipoleRep}b),
and the structure of water differs from that at zero field. 
Once this large polarization value is reached, additional polarization can be achieved through electronic
polarization. 
Note that this type of polarization is absent in rigid, fixed point charge models like SPC/E. 
We first quantify the changes in water structure induced by uniform electric fields through the site-site pair radial distribution functions (RDFs), $g(r)$, Fig.~\ref{fig:RDF}.
In the linear regime, the RDFs remain unchanged as the field is increased.
At high fields, the most significant changes are found in the O-O RDF,
in which the peaks sharpen and the minima deepen. 
There are small changes in the third and fourth peaks of the H-H RDF, and
the O-H RDF is essentially unchanged.
Importantly, the first, intramolecular peaks in the H-H and O-H RDFs remain unchanged, suggesting
that the bond lengths and angles are not altered by the field strengths studied here
and the increase in the dipole moment with field strength arises from electronic polarization.
This electronic response is captured in the SCFNN by its ability to describe the long range
interactions between the applied field and the molecular charge distribution. 
Because the changes in the pair structure at high field occur mainly at large distances,
they may be better quantified through X-ray scattering structure functions, $S(q)$, shown in Fig.~\ref{fig:RDF}d.
In the linear regime, $S(q)$ is unchanged as the field strength is increased, as found for the RDFs.
In the nonlinear regime, however, the first two peaks in $S(q)$ increase and the first minimum decreases
as the field strength is increased. 
The first peak consists mainly of O-H and O-O correlations, while the second peak is dominated by O-O correlations~\cite{Skinner2014},
such that the changes in $S(q)$ are consistent with O-O correlations changing most significantly at high fields.
At all field strengths, the intramolecular peaks in $S(q)$ at high $q$ are unchanged, further suggesting that
the molecular geometry is unchanged as high fields are applied.
These changes in $S(q)$ clearly indicate an increase in intermolecular ordering at high field strengths. 
%

\begin{figure}[tb]
\begin{center}
\includegraphics[width=2.8in]{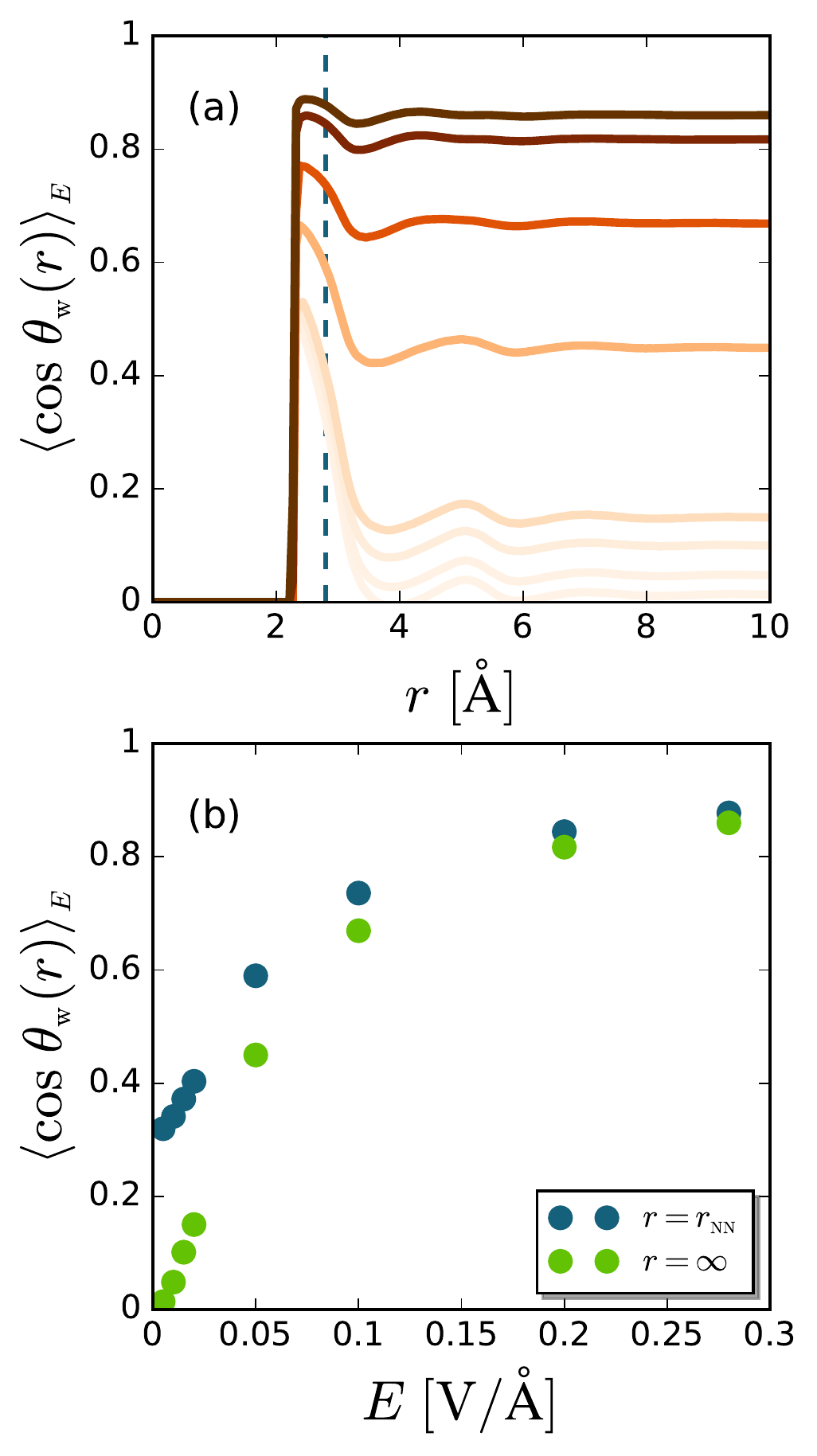}
\caption{\label{fig:costheta} (a) Cosine of the angle between the dipole moment vectors of two water molecules separated by a distance $r$ at different applied electrostatic fields. 
The darkness of the lines are proportional to the field strength.
The vertical dashed line indicates the location of the first peak in $g_{\rm OO}(r)$, which is used
to define the nearest neighbor distance $r_{\rm NN}$.
(b) Electric field dependence of the cosine of the angle between the dipole moment vectors of two water molecules
separated by a nearest neighbor distance, $r_{\rm NN}$, defined by the location of the first peak in the oxygen-oxygen RDF,
and two water molecules separated by large distances, $r=\infty$.}
\end{center}
\end{figure}

%
Although the RDFs and $S(q)$ show some increase in translational ordering of water,
the snapshots in Fig.~\ref{fig:DipoleRep} suggests that the changes in water structure are mainly in orientational ordering. 
To quantify the changes in orientational structure, we compute $\avg{\cos\theta_{\rm w}(r)}_E$,
where $\theta_{\rm w}(r)$ is the angle between the dipole moment vectors of two water molecules separated by a distance $r$, shown in Fig.~\ref{fig:costheta}a.
The orientational structure of water, as quantified by $\avg{\cos\theta_{\rm w}(r)}_E$ changes significantly with the strength of the applied electric field. 
At small fields, $\avg{\cos\theta_{\rm w}(r)}_E$ displays a large peak near 2.5~\AA \ that corresponds to dipolar ordering of neighboring water molecules. 
After this peak, $\avg{\cos\theta_{\rm w}(r)}_E$ exhibits small oscillations and tends to small values at large $r$; $\avg{\cos\theta_{\rm w}(r)}_E$ goes to zero at large distances in the absence of an applied field.
As \EE \ increases, the first peak in $\avg{\cos\theta_{\rm w}(r)}_E$ grows until the polarization begins to saturate, at which point the peak height changes very little with field. 
In addition, $\avg{\cos\theta_{\rm w}(r)}_E$ for large $r$ plateaus at higher values with large fields. 
This indicates long range ordering of water molecules in the presence of the electric field. 
Moreover, at the highest fields studied, $\avg{\cos\theta_{\rm w}(r)}_E$ is comparable in magnitude at all distances, suggesting that water is approaching maximal orientational order.
This highly ordered state is responsible for the plateau in the polarization and the saturation of the dielectric constant at high fields. 
The behavior of $\avg{\cos\theta_{\rm w}(r)}_E$ sheds light on the assumptions made in the Kirkwood-Booth Equation~\ref{eq:booth}.
At low field strength, $\avg{\cos\theta_{\rm w}(r)}_E$ exhibits a large peak in the first coordination shell, but is small after that,
lending support to Kirkwood's model for dipolar order in water~\cite{kirkwood1939dielectric,frohlich1948general,Booth1951}.
Furthermore, the value of $\avg{\cos\theta_{\rm w}(r)}_E$ at a typical nearest neighbor distance, $r_{\rm NN}$, is close
to the value of 1/3 used in Kirkwood's model~\cite{kirkwood1939dielectric,bernal1933theory},
where we have defined $r_{\rm NN}$ by the location of the first peak in the O-O RDF.
However, as the field strength increases into the non-linear regime, the correlations between nearest neighbors increase and
those beyond the first shell become significant with increasing field strength, Fig.~\ref{fig:costheta}b.
These field-dependent changes in pair structure are ignored in the simplistic Kirkwood-Booth model,
and inclusion of these effects may improve its accuracy.
%


\section{Conclusion}

We have applied the SCFNN model of water to dielectric saturation in water. 
For large uniform electric fields with magnitudes of approximately 0.05~V/\AA \ or larger,
water responds non-linearly to the applied field and the induced polarization starts to plateau.
This plateau arises from maximal reorientation of water molecules.
The increased alignment of water dipoles restricts their fluctuations and consequently lowers the dielectric constant.
This dielectric saturation can be described reasonably well with the Kirkwood-Booth theory~\cite{Booth1951,kirkwood1939dielectric,frohlich1948general},
and we have examined some of the key assumptions in this model.
Despite the saturation of the orientational structure of water, the electronic structure continues to respond
at the highest field strengths studied here, evidenced by the dipole moment increasing in magnitude, albeit non-linearly,
at high fields.
Despite the non-linear increase of molecular dipole moments due to electronic polarization, fluctuations
of the molecular dipoles remain Gaussian at all fields studied here.
The SCFNN model was not trained on configurations or fields typical of dielectric saturation.
Instead, non-linear response emerges naturally within the SCFNN framework because
the model learns the long range response responsible for dielectric screening and how
this response impacts the short range structure and interactions in water. 
Our results highlight that the SCFNN can be transferable, something that is beyond
the reach of many machine learning-based models.
We expect that the transferability of the current SCFNN model is largely limited to the long range response,
while the short range interactions will need to be retrained when additional local interactions
are introduced, such as those between water and ions.
However, we anticipate that the SCFNN idea of focusing on physical origins and length scales of molecular interactions,
and appropriately adapting the resulting neural network structure,
could enable the development of more transferable neural network models.

\begin{acknowledgments}
We acknowledge the Office of Advanced Research Computing (OARC) at Rutgers, The State University of New Jersey for providing access to the Amarel cluster and associated research computing resources. 
This work used Anvil at Purdue through allocation CHE210081 from the Advanced Cyberinfrastructure Coordination Ecosystem: Services \& Support (ACCESS) program, which is supported by National Science Foundation grant number 2005632.
\end{acknowledgments}


\bibliography{references}

\end{document}